\documentclass[10pt,prd,onecolumn,nofootinbib,
eqsecnum,byrevtex,amsfonts,amssymb,amsmath,draft]{revtex4}
\usepackage[dvips]{graphicx}
\usepackage[scriptsize,bf]{caption}
\usepackage{xspace}
\newsavebox{\oldtext}
\newcommand{\const}{\mathrm{const.}}

\newcommand{\br}[1]{\left(#1\right)}
\newcommand{\ud}[1]{\mathrm{d}#1}
\renewcommand{\exp}[1]{e^{#1}}
\newcommand{\sq}[1]{\left[#1\right]}

\newcommand{\abs}[1]{\left|#1\right|}

\renewcommand{\vec}[1]{\boldsymbol{#1}}
  
  \newcommand{\dd}[1]{\dot{#1}\dot{#1}}

\begin{document}
\title{On the nonuniqueness of free motion of the fundamental relativistic rotator}
\author{{\L}ukasz \surname{Bratek}}
\email[]{Lukasz.Bratek@ifj.edu.pl}
\affiliation{Henryk Niewodnicza{\'n}ski Institute of Nuclear Physics, \\
Polish Academy of Sciences, Radzikowskego 152, PL-31342 Krak{\'o}w, Poland}

\date{\today}
\begin{abstract}

  Consider a class of relativistic rotators described by position and a single null direction. Such a rotator is called fundamental if both its Casimir invariants are intrinsic dimensional parameters independent of arbitrary constants of motion. As shown by Staruszkiewicz, only one rotator with this property exists (its partner with similar property can be excluded on physical grounds).

  We obtain  a general solution to equations of free motion of the fundamental relativistic rotator in a covariant manner.
  Surprisingly, this motion is not entirely determined by initial conditions but depends on one arbitrary function of time, which specifies rotation of the null direction in the centre of momentum frame. This arbitrariness is in manifest contradiction with classical determinism. In this sense the \underline{isolated} fundamental relativistic rotator is pathological
   as a dynamical system.
 To understand why this is so, we study the necessary condition for the uniqueness of the related Cauchy problem.
  It turns out that the fundamental relativistic rotator (together with its partner) can be uniquely characterized  by violating this condition in the considered class of rotators.
  \end{abstract}
  \maketitle

\section{Introduction}
Fundamental relativistic rotator is a dynamical system described by a single null direction $k^{\mu}$ associated with position $x^{\mu}$ and moving  on extremals of the following Hamiltons' action \cite{bib:astar1}
\begin{equation}\label{eq:action}S=-m\int{\ud{\tau}\sqrt{\dot{x}\dot{x}}
 \sqrt{1+\sqrt{-\ell^2\frac{\dot{k}\dot{k}}{\br{k\dot{x}}^2}}}}.\end{equation}
  A dot denotes differentiation with respect to an arbitrary parameter $\tau$. In a given reference frame this rotator has $5$ degrees of freedom: $3$ for the spatial position plus $2$ for the null direction.\footnote{We remind that a null direction is a class of all  collinear null vectors.} The Casimir invariants of the Poincar\'{e} group   are constructed from the Noether constants of motion $P_{\mu}$ and $M_{\mu\nu}$, and  read\footnote{$ab\equiv{}a^0b^0-a^1b^1-a^2b^2-a^3b^3$,  $\epsilon^{\mu\nu\alpha\beta}$ is a completely antisymmetric pseudotensor for which $\epsilon^{0123}=1$.
  } $$P_{\mu}P^{\mu}=m^2,\qquad  W_{\mu}W^{\mu}=-\frac{1}{4}m^4\ell^2,$$ where $W^{\mu}$ is the Pauli-Luba\'{n}ski spin-vector defined as $W^{\mu}=-\frac{1}{2}\epsilon^{\mu\alpha\beta\gamma}M_{\alpha\beta}P_{\gamma}$.

As stated in \cite{bib:astar1}, the Hamiltons' action (\ref{eq:action})  is the only\footnote{\label{ft:plusminus}More precisely, also an action with $\sqrt{1-\sqrt{Q}}$ in place of $\sqrt{1+\sqrt{Q}}$ is possible, but we do not consider it since expression $1-\sqrt{Q}$ might become negative for sufficiently rapid rotation of the null direction in the centre of momentum frame.} relativistically invariant action composed of position, null direction, and their first derivatives, for which numerical values of both Casimir invariants are not  arbitrary constants of motion but are fixed by  intrinsic dimensional parameters of mass $m$ and length $\ell$. For that reason the rotator is called fundamental. For completeness we give below the relevant calculations since they were not included in the original paper.

   \medskip\noindent
Consider a class of relativistic rotators defined by the following Hamilton's action\footnote{Actions (\ref{eq:action}), (\ref{eq:act_f}) and function $Q$ have two spurious degrees of freedom: they are reparametrization invariant and also invariant under multiplication of null vector $k$ by arbitrary function (therefore we say they depend  on a null direction rather than on a null vector).}
\begin{equation}\label{eq:act_f} S=-m\int\ud{\tau} \sqrt{\dd{x}}f(Q),\qquad Q\equiv{-\ell^2\frac{\dd{k}}{\br{k\dot{x}}^2}}, \qquad kk=0.\end{equation}
The momenta canonically conjugated with $x$ and $k$ are, respectively,
  $$P\equiv-\frac{\partial{}L}{\partial{}\dot{x}}=\frac{mf(Q)}{\sqrt{\dd{x}}}\dot{x}
  -2mQf'(Q)\frac{\sqrt{\dd{x}}}{k\dot{x}}k\qquad \mathrm{and} \qquad \Pi\equiv-\frac{\partial{}L}{\partial{}\dot{k}}=2mQf'(Q)\frac{\sqrt{\dd{x}}}{\dd{k}}\dot{k}.$$
  We infer from the invariance of the Hamilton's action with respect to Poincar{\'e} transformations, that  the momentum vector  $P$ and the angular momentum tensor $M$ are conserved for solutions. Indeed, since a general variation of the Lagrangian reads\footnote{For the purpose of this section it suffices  to keep in mind that $kk=0$. However, in order to find equations of motion in a covariant form, one must add to the Hamilton's action an appropriate term with a Lagrange multiplier (which we shall do later).}
  $$\delta{L}=-\frac{d}{d\tau}\br{P\delta{x}+\Pi\delta{{k}}}+
  \dot{P}\delta{x}+\br{\dot{\Pi}+2mQf'(Q)\frac{\sqrt{\dd{x}}}{k\dot{x}}\dot{x}}\delta{k},
 $$  then, for infinitesimal global space-time translations $\epsilon$ and rotations $\Omega$ of solutions, we have
 \begin{eqnarray*}\br{\delta{}x=\epsilon=\const,\quad \delta{k}=0}\quad\Rightarrow\quad P=\const\\
 \br{\delta{x}^{\mu}=\Omega^{\mu}_{\phantom{\mu}\nu}x^{\nu},\quad \delta{k}^{\mu}=\Omega^{\mu}_{\phantom{\mu}\nu}k^{\nu},\quad \Omega_{\mu\nu}=\const,\quad{\Omega_{(\mu\nu)}=0}}\quad\Rightarrow\quad M_{\mu\nu}\equiv{}x_{\mu}P_{\nu}-x_{\nu}P_{\mu}+k_{\mu}\Pi_{\nu}-k_{\nu}\Pi_{\mu}=\const\end{eqnarray*}
  Casimir invariants of the Poincar\'{e} group are $P_{\mu}P^{\mu}$ and $W_{\mu}W^{\mu}$, where
 $W$ is the Pauli-Luba\'{n}ski (spacelike) vector defined earlier,
 hence
 $$PP=m^2\br{f^2(Q)-4Qf(Q)f'(Q)},\qquad WW=-4m^4\frac{Q^2f^2(Q)f'^2(Q)}{\br{\dd{k}}^2}
 \left|\begin{array}{ccc}
 kk& k\dot{k}& k\dot{x}\\
 \dot{k}k& \dot{k}\dot{k}& \dot{k}\dot{x}\\
 \dot{x}k& \dot{x}\dot{k}& \dot{x}\dot{x}\\
  \end{array}\right|
  =-4m^4\ell^2Qf^2(Q)f'^2(Q).$$
  By requiring that $PP=m^2$, we obtain $f(Q)=\pm\sqrt{1\pm a^2\sqrt{Q}}$. We may set the unimportant integration constant as $a^2=1$ since it only redefines parameter $\ell$.
  Note the remarkable fact that only for this particular $f(Q)$ the second Casimir invariant is also independent on $Q$, since then $WW=-\frac{1}{4}m^4\ell^2$. From physical reasons\footnote{Hamilton's action must have appropriate overall sign. See also footnote \ref{ft:plusminus}.} we choose $f(Q)=+\sqrt{1+\sqrt{Q}}$.

  The existence of the fundamental relativistic rotator is indeed very remarkable, since there is no apparent reason for two differential equations, originating from different physics, to  have a common solution. Only the fundamental relativistic rotator has the particular property that its mass and spin are intrinsic dimensional parameters. For any other $f$, spin and mass of the corresponding rotator     can not be simultaneously independent on the initial conditions.  In this sense the fundamental relativistic rotator is as fundamental as Dirac's electron.

\medskip

Unfortunately, as we shall see in the next section, the general solution of the fundamental relativistic rotator in free motion contains one arbitrary function of time, which describes angular velocity of the null direction in the centre of momentum frame. For a  physical dynamical system such ambiguity should not take place, since otherwise the system might accelerate or decelerate at will without apparent cause, which in turn, would mean lack of classical determinism. The determinism is closely related to uniqueness of the initial value problem that in mathematical physics is a postulate which can not be abandoned.
 Therefore, to understand why this arbitrary function is present, we analyze in section (\ref{sec:cauchy}) the question of solvability of the Cauchy problem for the fundamental relativistic rotator. As we shall see, this is the particular choice of the Hamilton's action of the fundamental  rotator  which makes the problem unsolvable. This serious deficiency of the fundamental rotator disappears already as a result of arbitrarily small deformation of function
$$\sqrt{1+\sqrt{Q}}, \qquad Q=-\ell^2\frac{\dot{k}\dot{k}}{\br{k\dot{x}}^2}$$
in action (\ref{eq:action}). The non-uniqueness has therefore nothing to do with the number of degrees of freedom and with the symmetries of $Q$, but is inherent in the fundamental relativistic rotator (and its partner with $f(Q)=\sqrt{1-\sqrt{Q}}$).

\section{Construction of solutions}
In what follows we shall find covariant form of solutions to the equations of motion resulting from Hamiltons' action (\ref{eq:action}).

\noindent
The generalized momentum $p^{\mu}$ corresponding to spacetime coordinates ${x}^{\mu}$ is
\begin{equation}\label{eq:p}
{p^{\mu}}\equiv-\frac{\partial{L}}{\partial{\dot{x}^{\mu}}}=m\br{\exp{\Psi}
u^{\mu}-\sinh\br{\Psi}\frac{k^{\mu}}{ku}},\qquad\mathrm{where}\quad u^{\mu}\equiv\frac{\dot{x}^{\mu}}{\sqrt{\dd{x}}}\quad \mathrm{and}\quad \exp{2\Psi}\equiv{\sqrt{-
\ell^2\frac{\dd{k}}{\br{\dot{x}k}^2}}+1}\quad (\Psi\geqslant0).
\end{equation}
It follows that $p_{\mu}p^{\mu}=m^2$. The auxiliary function $\Psi$ defined in (\ref{eq:p}) allows not only for concise notation of complicated expressions, but it has also a definite meaning. Namely, $\Psi$ is the hyperbolic angle between momentum $p^{\mu}$ and world velocity $u^{\mu}$, $pu=m\cosh{\Psi}$. Later, we shall come to the conclusion that $\frac{2}{\ell}\tanh{\Psi}$ is the angular velocity with which $k^{\mu}$ moves on the unit sphere of null directions in the centre of momentum frame.

The generalized momentum corresponding to null direction ${k}^{\mu}$ can be now concisely written as $$\pi_{\mu}\equiv-\frac{\partial{L}}{\partial{\dot{k}^{\mu}}}=m\,\sqrt{\dd{x}}\,
\sinh\br{\Psi}\frac{\dot{k}_{\mu}}{{\dd{k}}}=-\frac{m^2\ell}{2 pk}\,\frac{\dot{k}_{\mu}}{\sqrt{-\dd{k}}},$$
where we have used  the identity $2pk\sqrt{\dd{x}}\sinh{\Psi}=\ell{}m\sqrt{-\dd{k}}$ resulting from (\ref{eq:p}).

A convenient way of deriving the equation of motion for $k^{\mu}$, without the need for introducing internal coordinates on the cone $kk=0$,  is by finding a conditional extremum of functional (\ref{eq:action}) subject to the condition $kk=0$. This is a standard variational problem with subsidiary conditions. The stationary value of functional (\ref{eq:action}) with the condition $kk=0$, can be found by supplementing the functional with an additional term $\int\ud{\tau}(-)\Lambda(\tau)kk$ containing a Lagrange multiplier $\Lambda(\tau)$.  By varying such extended action with respect to $\Lambda$, we restore our condition $kk=0$, whereas a variation with respect to vector $k$, yields the following equation $\dot{\pi}_{\mu}+\partial_{{k}^{\mu}}L-2\Lambda k_{\mu}=0$. By contracting it with vector $p^{\mu}$, we  find the unknown function $\Lambda(\tau)$, and hence, obtain the desired equation of motion for $k$
\begin{equation}\label{eq:tensor}\br{\dot{\pi}_{\nu}+
 \frac{\partial{}L}{\partial{k}^{\nu}}}\br{\delta^{\nu}_{\phantom{\nu}\mu}-
 \frac{p^{\nu}k_{\mu}}{pk}}=0,\qquad  kk=0.\end{equation}
We have not written this  complicated equation explicitly, as it can be considerably simplified  and recast in a form having a very clear geometrical meaning.

First, note that a null vector $k^{\mu}$ can always be written as $k^{\mu}=h\br{m^{-1}p^{\mu}+n^{\mu}}$, where $n^{\mu}$ is a unit spacelike vector orthogonal to timelike vector $p^{\mu}$, and $h=m^{-1}p^{\mu}k_{\mu}$.
  Secondly, for describing a space-like curve $n^{\mu}(\tau)$, it is more natural to regard its arc length $$\phi\br{\tau}=
\int\ud{\tau}\sqrt{-\dd{n}},\qquad nn=-1,\quad np=0,
$$ as an independent variable, rather than any other. Furthermore, the momentum $p^{\mu}$ is conserved, $p^{\mu}(\tau)=P^{\mu}$, where $P^{\mu}$ is a constant vector such that $P_{\mu}P^{\mu}=m^2$.
Now, making use of these observations,  equation (\ref{eq:tensor}) can be reduced (up to unimportant $h$-dependent factor) to the following equation for $n$
 \begin{equation}\label{eq:n}\frac{\ud{}^2n^{\mu}}{\ud{}\phi^2}+n^{\mu}=0, \qquad nn=-1,\quad nP=0,\end{equation} which is nothing but the equation for great circles on a unit sphere in the subspace orthogonal to $P^{\mu}$ (then $\phi(\tau)$ is the angle).
Expressed in terms of $n^{\mu}(\phi)$, the Pauli-Luba\'{n}ski spin-vector reads
$$W^{\mu}=\frac{1}{2}m\ell\epsilon^{\mu\alpha\beta\gamma}n_{\alpha}
\frac{\ud{n_{\beta}}}{\ud{\phi}}P_{\gamma}.$$
This constant vector is orthogonal to the plane spanned by $n^{\mu}$ and $\frac{\ud{n^{\mu}}}{\ud{\phi}}$, thus, together with $P^{\mu}$,  it can be used  to construct solutions.

A parametric description of a specific circle from the family of solutions, can be visualized as a continuous action of an elliptic Lorentz transformation upon some fixed unit spatial vector $N^{\mu}$ orthogonal to $W^{\mu}$ and $P^{\mu}$. Such a transformation must leave invariant two null directions $K_{\pm}^{\mu}=\frac{1}{\sqrt{2}}\br{\frac{P^{\mu}}{m}\pm\frac{W^{\mu}}{
\frac{1}{2}m^2\ell}}$. Parameterized by elliptic angle $\phi$, the general solution for $n^{\mu}(\phi)$ is thus easily found to be
$$n^{\mu}\br{\phi}
=N^{\mu}\cos{\phi}-\frac{\epsilon^{\mu\nu\alpha\beta}N_{\nu}W_{\alpha}P_{\beta}}{
\frac{1}{2}m^3\ell}\sin{\phi}, \qquad NN=-1,\quad NW=0,\quad NP=0,$$
which is indeed a solution to equation (\ref{eq:n}). As was to be anticipated from the independence of Hamilton's action (\ref{eq:action}) upon scaling of the null vector $k$ by arbitrary function, there is no constraint imposed on function $h$   by the equations of motion, thus, without lose of generality, we may set $h\equiv1$. Finally, the corresponding null direction and spacetime position   can be now found from
\begin{equation}\label{eq:sol}k^{\mu}={n^{\mu}+\frac{P^{\mu}}{m}},\qquad \frac{{\dot{x}}^{\mu}}{\sqrt{\dd{x}}}=\frac{P^{\mu}}{m}\cdot\cosh{\Psi}+
n^{\mu}\cdot\sinh{\Psi}.\end{equation} The second equation comes from Noether integral (\ref{eq:p}).

\medskip\noindent
It is rather astonishing to find out that
function $\Psi$ is not determined by the equations of motion, but
it is best that we postpone this important issue until next section. Now we only shortly explain  the physical meaning of this function.

  Function $\Psi$ is related to the time dependence of rotation of the null direction in the centre of momentum frame. The proper time in this frame increases by $\ud{t}=\br{m^{-1}P_{\mu}}\dot{x}^{\mu}\ud{\tau}$ with every infinitesimal displacement $\ud{x}^{\mu}=\dot{x}^{\mu}\ud{\tau}$ of the rotator. It follows from equation (\ref{eq:sol})  that $\sqrt{\dot{x}\dot{x}}\,\sinh{\Psi}=-n_{\mu}\dot{x}^{\mu}$ and, in conjunction with the definition of $\Psi$ in equation (\ref{eq:p}),    $2\sqrt{\dot{x}\dot{x}}\,\sinh{\Psi}\ud{\tau}${}$=\ell\sqrt{-\dot{n}\dot{n}}\,
 \ud{\tau}${}$\equiv\ell\,|\dot{\phi}|\,\ud{\tau}$, where $\dot{\phi}\,\ud{\tau}$ is the change in the angular position of the null direction as observed in this frame. Therefore, the angular speed of the rotator  measured  in the center of momentum frame is
$$\abs{\frac{\ud{\phi}}{\ud{t}}}=
-\frac{2m}{\ell}\cdot\frac{n_{\mu}u^{\mu}}{P_{\mu}u^{\mu}}=\frac{2}{\ell}\tanh{\Psi}
<\frac{2}{\ell}.$$
To solve equation (\ref{eq:sol}) we may choose the arbitrary parameter $\tau$ so as $\tau\equiv{}t$ (hereafter $\dot{\square{}}\equiv\frac{\ud\square}{\ud{t}}$).\footnote{Hamilton's action (\ref{eq:action}) is reparametrization-invariant, thus $\tau$ can be an arbitrary parameter such that ${\frac{\ud{x^0}}{\ud{\tau}}}$ is continuous and everywhere nonzero.} Now, on account of the earlier definition of $t$, $\ud{t}=\br{m^{-1}P_{\mu}}\dot{x}^{\mu}\ud{\tau}$, we have $P_{\mu}\dot{x}^{\mu}\equiv{}m$, or equivalently, $\cosh{\Psi}=\br{\dd{x}}^{-1/2}$. Hence, $\dot{x}^{\mu}\ud{t}=\frac{P^{\mu}}{m}\ud{t}+n^{\mu}\tanh{\Psi}\ud{t}=
\frac{P^{\mu}}{m}\ud{t}+\frac{\ell}{2}n^{\mu}\ud{\phi}$, and finally, integration gives $x^{\mu}(t)$.

\medskip
As follows from the above derivation, the equations of free motion of the fundamental relativistic rotator  can be solved exactly. Here we use vector $r^{\mu}$ defined by  $\dot{r}^{\mu}(t)=n^{\mu}(\phi(t))\dot{\phi}(t)$ rather than $n^{\mu}$,  then $|{\dot{\phi}(t)}|=\sqrt{-\dot{r}(t)\dot{r}(t)}$.
\begin{quote}
\textbf{General solution.} Free motion of the dynamical system  defined by Hamilton's action (\ref{eq:action})
has the following, relativistically invariant, parametric description
$$x^{\mu}(t)=\frac{P^{\mu}}{m}t+\frac{\ell}{2}r^{\mu}(t)+x^{\mu}(0),\qquad \mathrm{and}\qquad k^{\mu}\br{t}=\frac{P^{\mu}}{m}+\frac{\dot{r}^{\mu}(t)}{{\sqrt{-\dot{r}(t)\dot{r}(t)}}},$$
where
$$r^{\mu}(t)=N^{\mu}\sin{\phi(t)}+
\frac{\epsilon^{\mu\nu\alpha\beta}N_{\nu}W_{\alpha}P_{\beta}}{
\frac{1}{2}m^3\ell}\cos{\phi(t)}.$$

Constant vectors $P^{\mu}$, $W^{\mu}$ and $N^{\mu}$ satisfy the following conditions
$$PP=m^2,\qquad WW=-\frac{1}{4}m^4\ell^2,\quad WP=0,\qquad NN=-1,\quad NW=0,\quad NP=0.$$ $P^{\mu}$ is the (conserved) momentum of the centre of momentum frame, $t$ is the proper time in this frame, and $W^{\mu}$ is the (conserved) intrinsic angular momentum (spin) of the rotator.

Function $\phi(t)$ describes the angular position of the "pointer" $k^{\mu}(t)$ in the centre of momentum frame.
{\emph{For it is not determined by the equations of motion, this function is arbitrary.}}
More precisely, it may
 be any function for which $0<|{\dot{\phi}(t)}|<\frac{\ell}{2}$, that is, $\dot{\phi}(t)$ must be always nonzero.

\end{quote}
 Hamiltons' action (\ref{eq:action}) evaluated for this general solution reads
$$S(t)=S(0)-m\,t-\frac{m\,\ell}{2}\,\phi(t), \qquad (\mathrm{if\ } \dot{\phi}(t)>0).$$
The first term in $S$ is the ordinary contribution from the proper time of the centre of momentum frame, and the second is the corresponding contribution from the intrinsic spin of the rotator.

\section{On the  Cauchy problem for the fundamental relativistic rotator}\label{sec:cauchy}

The necessary condition for the existence of Hamiltonian mechanics for a dynamical system
described by a Lagrangian $L(v,q)$, is that for fixed $q$ the set of equations $p(v,q)=\frac{\partial{}L}{\partial{}v}(q,v)$ defining momenta $p$, should be a diffeomorphism of spaces of momenta $p$ and of velocities $v$. In particular, the set of equations should be uniquely solvable  for velocities $v=v(q,p)$. This is possible provided $$\det\sq{\frac{\partial{}^2L}{\partial{}\dot{q}^i\partial\dot{q}^j}}\ne0,$$ otherwise the Legendre transform leading from Lagrangian to Hamiltonian is not well defined.

The above condition can be equivalently viewed as  necessary for  unique dependence of accelerations on the initial data. The
Euler-Lagrange equations for $L$ can be recast in the general form
$$\frac{\partial{}^2L}{\partial{}\dot{q}^i\partial\dot{q}^j}\ddot{q}^j=Z(q,\dot{q},t),$$ with some function $Z$. Therefore, the vanishing of the determinant would not only mean that accelerations could not be algebraically determined from the positions $q$ and their derivatives, but also that equations of motion could not be reduced to the canonical form  $\dot{y}=F(y,t)$, where $y=(q,\dot{q})$, for which general results on the existence and uniqueness of solutions of ordinary differential equations were obtained.

\subsection{A proof that  the fundamental relativistic rotator (and its partner)\\is uniquely characterized by the condition $\det\sq{\frac{\partial{}^2L}{\partial{}v^i\partial{}v^j}}\equiv0$. }

As we have seen, solvability of the Cauchy problem for a class of relativistic rotators defined by the general action
$$S=-m\int\ud{\tau} \sqrt{\dd{x}}f\br{-\ell^2\frac{\dd{k}}{\br{k\dot{x}}^2}},$$  with $f$ being arbitrary function,
 can be examined by answering the question whether or not the determinant of the matrix of second derivatives of the resulting Lagrangian
vanishes or not.

In what follows we shall show that the determinant is zero only for Hamilton's action  of the fundamental relativistic rotator (\ref{eq:action}). This result means that the equations of motion  can not be solved with the help of Picard's method. The motion of the rotator may be therefore indeterministic, which explains the presence of arbitrary function in our general solution found in the previous section.

\medskip
For our purpose it suffices to consider some particular map adapted to constraints. We use Cartesian map for the space-time position and spherical angels for the null direction, and the arbitrary parameter is chosen so as $\tau\equiv{}x^0$, thus\footnote{We note that the presence of arbitrary function of time in the general solution has nothing to do with reparametrization-invariance of the Hamilton's action. Secondly,
we have fixed here the arbitrary parameter $\tau=x^0$ and internal coordinates to eliminate gauge functions. It should be clear that we do not lose generality of our proof by choosing the particular map. By that we are left only with $5$ physical degrees of freedom.}
$$x^0(t)=\ell{}t,\qquad \vec{x}(t)=\ell\sq{X^1(t),X^2(t),X^3(t)},\qquad k^0(t)=1, \qquad \vec{k}(t)=\sq{\sin{\theta(t)}\cos{\phi(t)},\sin{\theta(t)}\sin{\phi(t)},\cos{\theta(t)}}.$$
In this parametrization the Lagrangian is proportional to function
$$\mathcal{L}\br{V,W}=\sqrt{1-V^TV}f(Q),\qquad Q=\frac{W^TW}{\br{1-N^TV}^2},$$
where
$$V=\left(\begin{array}{c}\dot{X}^1(t)\\ \dot{X}^2(t)\\ \dot{X}^3(t)\end{array}\right), \qquad W=\left(\begin{array}{c}\dot{\theta}(t)\\ \dot{\phi}(t)\sin{\theta(t)}\end{array}\right),
\qquad N=\left(\begin{array}{c}\sin{\theta(t)}\cos{\phi(t)}\\ \sin{\theta(t)}\sin{\phi(t)}\\ \cos{\theta(t)}\end{array}\right).$$
The determinant of the matrix of  second derivatives of the Lagrangian with respect to
velocities $\dot{X}^1$, $\dot{X}^2$, $\dot{X}^3$, $\dot{\theta}$, $\dot{\phi}$
 is proportional to the determinant of the following symmetric matrix of size $5\times5$
$$H=\left[\begin{array}{cc}A&B\\ B^T&C\end{array}\right],$$
where\footnote{
Note that $A=A^T$ and $C=C^T$ are of size $2\times{2}$ and $3\times{3}$, but $B$ and $B^T$ are matrices of different shape, of size $2\times{3}$ and $3\times{2}$, respectively.
We remind also the obvious thing that the order of multiplication is important, e.g $WV^T$ is a rectangular matrix with $2$ verses and $3$ columns, $NV^T$ is a $3\times3$ square matrix, while $N^TV=V^TN$ is a scalar product of column vectors $N$ and $V$. }
\begin{eqnarray*}
A&=&
2Qf'(Q)\frac{\sqrt{1-V^TV}}{W^T{W}}\br{I+
2\frac{Qf''(Q)}{f'(Q)}\frac{WW^T}{W^T{W}}}
\\
B&=&
2Qf'(Q)\frac{\sqrt{1-V^TV}}{W^T{W}}
\br{2\sq{1+\frac{Qf''(Q)}{f'(Q)}}\frac{WN^T}{1-N^T{V}}-
\frac{WV^T}{1-V^T{V}} },
\\
C&=&
-\frac{f(Q)}{\sqrt{1-V^TV}}\br{I+\frac{VV^T}{1-V^TV}
+2\frac{Q\,f'(Q)}{f(Q)}\sq{\frac{NV^T+VN^T}{1-N^T{V}}-\br{3+2\frac{Qf''(Q)}{f'(Q)}}
\frac{1-V^T{V}}{\br{1-N^T{V}}^2}NN^T}}
\end{eqnarray*}
 By $I$ we denote the identity matrices of appropriate size. The elements of matrices $A$, $B$ and $C$ are numerically equal to the respective second derivatives $$A^{i'}_{j'}\hat{=}\frac{\partial^2\mathcal{L}}{\partial{W^{i'}}\partial{W^{j'}}},
\qquad B^{i'}_{j}\hat{=}{\frac{\partial^2\mathcal{L}}{\partial{W^{i'}}\partial{V^{j}}}}
={\frac{\partial^2\mathcal{L}}{\partial{V^{j}}\partial{W^{i'}}}}\hat{=}\br{B^{T}}^{j}_{i'},
\qquad C^{i}_{j}\hat{=}\frac{\partial^2\mathcal{L}}{\partial{V^{i}}\partial{V^{j}}},$$
Due to the structure of matrix $H$  the task of calculating its determinant simplifies significantly.
First we employ the following identity
$$\left[\begin{array}{cc}A&B\\ B^T&C\end{array}\right]=\left[\begin{array}{cc}A&0\\ B^T&I\end{array}\right]\cdot\left[\begin{array}{cc}I&A^{-1}B\\ 0&C-B^TA^{-1}B\end{array}\right],$$ holding for a block matrix composed of  matrices of mutually compatible dimensions. Hence, we conclude that $$\det(H)=\det(A)\det(C-B^TA^{-1}B).$$
By applying Sylvester's determinant theorem\footnote{In general, Sylvester's  theorem states that $\det(I_{{m\times{}m}} + RS) = \det(I_{{n\times{}n}} + SR)$ for matrices $R$ and $S$ of size $m\times{}n$ and $n\times{}m$, respectively, where $I_{{m\times{}m}}$ and $I_{{n\times{}n}}$ are unit matrices. In particular, for column vectors $a$ and $b$ of size $n$ we have $\det(I_n + ab^T) = \det(I_1 + b^Ta)=1+b^Ta=1+a^Tb$. } we can easily calculate $\det(A)$
$$\br{2Qf'(Q)\frac{\sqrt{1-V^TV}}{W^T{W}}}^{-2}\det(A)=\det\br{I+
2\frac{Qf''(Q)}{f'(Q)}\frac{WW^T}{W^T{W}}}=1+2\frac{Qf''(Q)}{f'(Q)}\frac{W^T{W}}{W^T{W}}=
1+2\frac{Qf''(Q)}{f'(Q)}.
$$
The inverse of $A$ can also be easily found by supposing that $A^{-1}=a\br{I+b WW^T}$ with $a$ i $b$ to be determined from the condition $A^{-1}A=I$. The result is
$$A^{-1}=\frac{W^T{W}}{2Qf'(Q)\sqrt{1-V^TV}}\br{I+
2\frac{Qf''(Q)}{f'(Q)}\frac{WW^T}{W^T{W}}}^{-1}=\frac{W^T{W}}{2Qf'(Q)\sqrt{1-V^TV}}\br{I-
2\frac{Qf''(Q)}{f'(Q)+2Qf''(Q)}\frac{WW^T}{W^T{W}}}.$$ By noting that $\frac{\br{WW^T}\br{WW^T}}{W^TW}=\frac{W\br{W^TW}W^T}{W^TW}=WW^T$, ect, we find  that
\begin{eqnarray*}
&B^TA^{-1}B=\frac{2Qf'(Q)}{1+\frac{2Qf''(Q)}{f'(Q)}}\sqrt{1-V^TV}
\br{2\br{1+\frac{Qf''(Q)}{f'(Q)}}\frac{N}{1-N^TV}-\frac{V}{1-V^TV}}
\br{2\br{1+\frac{Qf''(Q)}{f'(Q)}}\frac{N^T}{1-N^TV}-\frac{V^T}{1-V^TV}},&
\end{eqnarray*}
and next that
\begin{eqnarray*}
 &C-B^TA^{-1}B=-\frac{f(Q)}{\sqrt{1-V^TV}}
\left[
\br{I+\frac{VV^T}{1-V^TV}}+\frac{2Qf'(Q)}{f(Q)\br{1+\frac{2Qf''(Q)}{f'(Q)}}}
\frac{1-V^TV}{\br{1-N^TV}^2}
\br{N-\frac{1-N^TV}{1-V^TV}V}
\br{N^T-\frac{1-N^TV}{1-V^TV}V^T}\right].&
\end{eqnarray*}This is again a square matrix to which Sylvester's determinant theorem applies
\begin{eqnarray*}
& -\br{\frac{f(Q)}{\sqrt{1-V^TV}}}^{-3}\det{\br{C-B^TA^{-1}B}}=\\
&
\det\br{I+\frac{VV^T}{1-V^TV}}\br{
1+
\frac{2Qf'(Q)}{f(Q)\br{1+\frac{2Qf''(Q)}{f'(Q)}}}\frac{1-V^TV}{\br{1-N^TV}^2}
\br{N^T-\frac{1-N^TV}{1-V^TV}V^T}
\br{I-VV^T}
\br{N-\frac{1-N^TV}{1-V^TV}V}}=\\
&=\frac{1}{1-V^TV}\br{1+\frac{2Qf'(Q)}{f(Q)\br{1+2\frac{Qf''(Q)}{f'(Q)}}}},
\end{eqnarray*}
 where we have used the identity $\br{I+\frac{VV^T}{1+V^TV}}^{-1}=I-VV^T$ and performed a decomposition of matrix $C-B^TA^{-1}B$ similar to the previous one for matrix $H$.
Finally, on expressing $W^TW$ by $Q$ in the  formula for $\det{(A)}$ derived earlier, we obtain
$$\det{H}=-\frac{4f(Q)^3f'(Q)^2}{\br{1-N^TV}^4\br{1-V^TV}^{3/2}}\br{1+2Q\br{
\frac{f'(Q)}{f(Q)}+\frac{f''(Q)}{f'(Q)}}}.$$
The only nontrivial function $f(Q)$ for which the above determinant of $H$ is identically zero reads
$$f(Q)=c_1\sqrt{1+c_2\sqrt{Q}},$$
where $c_1$ and $c_2$ are integration constants, which can be absorbed by dimensional parameters of the model, thus we have only two physically distinct solutions $\sqrt{1\pm\sqrt{Q}}$. Solution $\sqrt{1+\sqrt{Q}}$ is precisely the function in the Lagrangian of the fundamental relativistic rotator!

\section{Conclusions}

The result of the previous section that function $f(Q)=\sqrt{1+\sqrt{Q}}$ (together with $f(Q)=\sqrt{1-\sqrt{Q}}$) in Hamilton's action (\ref{eq:act_f}) is uniquely determined by requiring that the determinant of a matrix of second derivatives of the respective Lagrangian with respect to velocities must be identically zero,  shows that
the case of the fundamental rotator is now even more striking than thought previously based on the requirement of paper \cite{bib:astar1} that $f(Q)$ should be such that  both Casimir invariants of the Poincar\'{e} group should be rather parameters and not arbitrary constants of motion. Now we see, that
there are three distinct differential equations for function $f$ (two of first order and one of second order) to which one arrives from different premises,
and which have common solution! This shows that the fundamental relativistic rotator is indeed a very particular dynamical system and somehow degenerated.

 The main result of this paper that the Cauchy problem for the fundamental relativistic rotator is not unique,  poses the question about existence  of  classical fundamental
systems, that is, such for which both Casimir invariants of the Poincar\'{e} group are parameters rather than arbitrary constants of motion. Although
free motion of the fundamental relativistic rotator is indeterministic,  which is rather  pathological, it seems that for the motion to be unique one needs interaction with external fields. The interaction term should be such to guarantee  non-singularity of a matrix of  second derivatives of the total Lagrangian with respect to velocities. Already, arbitrarily small deformation of function $f(Q)$ of the fundamental rotator removes this singularity (which shows the singularity has nothing to do with the number of degrees of freedom), however this deformation spoils the feature of being a fundamental system. One could therefore say that it looks as if Nature was trying to say us something very important, in particular, that isolated classical fundamental dynamical systems may be mathematically inconsistent, and that for consistency one would need appropriate interaction with other fields.

\acknowledgements
I would like to acknowledge Professor Andrzej Staruszkiewicz for his always invaluable and stimulating discussions.

\end{document}